\documentclass[10p,times,twocolumn]{article}

\usepackage{geometry}
\usepackage{amssymb}

\usepackage{amsmath}
\usepackage{subfigure} 
\usepackage[T1]{fontenc}

\usepackage{ marvosym }
\usepackage{tikz,pgfplots}

\usepackage[numbers,sort&compress]{natbib}
\usepackage[affil-it]{authblk}
\usepackage {amsmath}

\usepackage{tabularx}

\begin{document}

\title{\textbf{A three-dimensional scanning stereoscopic particle image velocimetry for large-volume liquid flows}}

\author{Paul van der Plas$^{1,2}$}
\author{{\color{black}Yutian He$^{3}$}}

\author{Varghese Mathai$^{3,4,2}$\thanks{Email: vmathai@umass.edu}}

\affil{\normalsize{$^{\textup{1}}$ Thales Systems BV, 7554 RR Hengelo, The Netherlands}}

\affil{\normalsize{$^{\textup{2}}$Physics of Fluids and Max Planck Center for Complex Fluid Dynamics, 7500 AE Enschede, The Netherlands}}

\affil{\normalsize{$^{\textup{3}}$ Department of Physics, University of Massachusetts Amherst, MA 01003, USA}}
\affil{\normalsize{$^{\textup{4}}$ Department of Mechanical and Industrial Engineering, University of Massachusetts Amherst, MA 01003, USA}}

\date{}

\maketitle

\paragraph{Abstract:} 
\textcolor{black}{We introduce a Scanning Stereoscopic Particle Image velocimetry~(SSPIV) system which incorporates a thin rotating prism to obtain three-dimensional velocity fields in a large cubical  measurement volume. The system makes use of an arrangement where a laser beam is first laterally deflected using an even-faced prism, and then passed through a cylindrical lens rod to obtain light sheets for illumination. This eliminates the need for large rotating mirrors or heavy prisms and facilitates relatively fast rotation speeds of up to 4800 rpm for the prism, yielding three-dimensional, three-component velocity field over a large depth at relatively high particle seeding density. About $10^6$ velocity vectors (3D 3C) are obtained in a single volumetric-scan of $100 \times 100 \times 100$ mm$^3$.
The method is demonstrated using two independent experiments, 1) a round jet at Reynolds number Re = 1000, and 2) the flow field around a freely rising sphere at a Reynolds number Re$_p$ $\approx$ 260, and the wake flow is compared with direct numerical simulations of a falling viscous droplet. The scanning gives time-resolved evolution of the large-scale vortical structures of the Re = 1000 jet, with spatial gradients captured fairly well. {\color{black}The three-dimensional flow field surrounding the refractive-index-matched rising sphere is shown to be steady and non-axisymmetric. The flow field and separation bubble are seen to be similar, but slightly shorter than that of the falling droplet case and the uniform flow past a fixed sphere case at a comparable Reynolds number.} }

\emph{Keywords: Stereoscopic PIV, Scanning, 3D3C, round jet, buoyant sphere, divergence, wake}

\section{Introduction}

Particle Image Velocimetry~(PIV) is a widely employed experimental technique that allows to resolve spatial variations of velocity in laboratory \cite{eggels1994pipeflowexp} and natural flows \cite{Hong2014snowfall}. In PIV the fluid motion is made visible by adding small tracer particles, and from the positions of these tracer particles at two instances in time, i.e. the particle displacement (the velocity field) is inferred~\cite{cite_piv1}. Since its introduction in the 1980's, PIV has evolved to become the dominant technique for velocimetry in the experimental fluid dynamics, and has been an important instrument for many advances in our understanding of turbulent and complex flows~\cite{cite_piv2}.
The single-camera PIV, is still the most widely used method for two-dimensional two-component (2D 2C), yielding velocity information on an illuminated plane.
While two-dimensional two-component (2D-2C) PIV \cite{adrian1991particle,brucker1995} is widely used for flow quantification in experimental fluid mechanics, in practice, the vast majority of fluid flows addressed are three-dimensional. 
An alternative is to use multiple cameras to view the same light sheet from two angles, i.e. by stereoscopic imaging. 
This gives three components of velocity on the illuminated two-dimensional sheet~(2D-3C), and thereby yielding six out of the nine spatial gradients of velocity~\cite{prasad2000stereoscopic,mathai2022fluid,kaufer2021stereoscopic,onoue2015large}.

In comparison, the tomographic PIV technique developed by {\citet{elsinga2006tomographic,scaranoTomoPIVMeasSciTech2012}} offers the possibility of obtaining fully resolved volumetric data, with all nine spatial velocity gradients captured. Since its advent, the method has been extensively employed to unravel the physics of a variety of fluid flows including boundary layers, cylinder wakes, jets, step flows, etc.
However, tomo-PIV requires a stronger laser source as compared to 2D PIV due to the additional requirement for volumetric illumination. Additionally, the seeding density of particles within this tomo-PIV volume is restricted, as compared to the seeding density possible in two-dimensional PIV. A compromise is often made by either reducing the particle seeding density at the expense of spatial resolution, or by illuminating a measurement volume that is shallow in the depth direction to allow sufficiently high seeding density\cite{elsinga2006tomographic,Scarano2009vortpatterns,Ghaemi2011hairpinvortices,Hain2008finitecylinder,Atkinson2011accuracytomo}. The latter implies that Tomo-PIV proves very successful in wall-bounded turbulent flows, since the wall normal distance to be resolved, i.e. the boundary layer height, is usually much smaller than the streamwise and spanwise directions.
Furthermore, the 3D reconstruction procedure is usually under-determined and hence needs to be solved iteratively, which makes it computationally expensive.

The Scanning Particle Image Velocimetry technique is an extension to the two-dimensional PIV where the light sheet is moved at a very fast speed normal to the illumination plane. 
When implemented with a single camera recording system~\cite{brucker1995,david2012time}, it can yield a three-dimensional two-component (3D 2C) velocity field, giving information on the two velocity components in a volume. With multiple cameras, the same can yield a 3D-3C velocity field, with all nine spatial velocity gradients resolved \cite{david2012time,casey2013scanning}. Since the scanning sheet is thin as compared to the volumetric illumination approaches, such as Tomo-PIV, the interference and scattering from in flows with a dispersed phase (bubbles or particles)is expected to be lower\cite{almeras2017experimental,ern2012wake,flapper2025multi}.  Scanning PIV systems have be implemented in literature by adopting various approaches. Majority of scanning systems have used oscillating mirrors for scanning~\cite{Gono2012buoyantjet,hori2004high,casey2013scanning,brucker1996a,Watanabe2015jetsspiv,lawson2014scanning,knutsen2017laser}.
These mirrors can be small in construction and precise, allowing for very good volumetric scan depths, which scale with the angular amplitude of the mirror. However, since the reflected beam from the scanning mirror diverges, they require converging optics. At very high scanning rates, oscillating mirrors are subjected to intense accelerations, which necessitate high precision optics. 

\citet{hoyer20053d} implemented a scanning particle tracking system by using a rotating prism in place of the oscillating mirror. Here, the laser beam was first  expanded into a sheet using a cylindrical lens, and then made to pass through the prism, which yielded a scan depth of about 20 mm. A few experiments \cite{hori2004high,knutsen2017laser,david2012time,bhat2018leading} have achieved large scan depths in the range 50-100 mm, however, with a relatively long duration of volume scan~[0.2, 0.33] s. In other cases, a compromise was usually made between the mirror's oscillation frequency and scan depth, in order for the  flow field to be considered frozen during the duration of each volume scan. The scanning speed (in mm/s) can be a useful metric to compare the different studies in literature. \citet{casey2013scanning} achieved one of the fastest scan speeds ($\sim$ 3750 mm/s) by smartly combining sub-volumes. In this experiment, however, the sub-volumes themselves were thick ($\sim$ 25 mm), and hence the velocity vectors were obtained iteratively using the MART algorithm~\cite{elsinga2006tomographic} of Tomo-PIV.

\newcolumntype{L}[1]{>{\raggedright\arraybackslash}p{#1}}
\newcolumntype{C}[1]{>{\centering\arraybackslash}p{#1}}
\newcolumntype{R}[1]{>{\raggedleft\arraybackslash}p{#1}}
\begin{table*} [!t]
\caption{A summary of the scanning parameters from literature. Here, x and y give the width and height of the camera field of view, respectively, with the product xy comparable to the laser sheet area. z is the depth of the volume scan. $\tau_{\mathcal{V}}$ is the time taken for a volume scan (ms). The scan rate gives the number of scans per second, and scan speed is the average speed at which the laser sheet traverses the scan volume, in mm/s.} 
\centering
\begin{tabular}{L{0.28\linewidth}R{0.06\linewidth}R{0.1\linewidth}R{0.08\linewidth}R{0.08\linewidth}R{0.09\linewidth}R{0.09\linewidth}}
\hline

Reference &x & y &  z & ${\tau_\mathcal{V}}$ & scan~rate & scan~speed\\
 & [mm] & [mm] & [mm] & [ms] & [Hz] & [m/s]\\
\hline\\
\citet{david2012time} & 185 & 140 & 83 & 100 & 10 & 0.83 \\ \\
\citet{casey2013scanning} & 100 & 100 & 75 & 20 & 50 & 3.75\\ \\
\citet{Gono2012buoyantjet} & 40 & 60 & 40 & 73 & 14 & 0.54 \\ \\
\citet{hori2004high} & 100 & 100 & 100 & 330 & 3 & 0.3 \\ \\
\citet{Watanabe2015jetsspiv} & 40 & 80 & 40 & 67 & 50 & 0.59\\ \\
\citet{lawson2014scanning} & 125 & 123 & 23 & 25 & 40 & 0.92\\ \\
\citet{hoyer20053d} & 20 & 20 & 20 & 20 & 50 & 1.0\\ \\
\citet{bhat2018leading} & 50 & 50 & 55 & 28 & 36 & 1.9\\ \\
\citet{albagnac2014three} & 300 & 300 & 35 & 200 & 5 & 0.17\\ \\

Present work & 100 & 100 & 100 & 10 & 100 & 10 \\
\hline
\hline
\end{tabular}
\label{table:lit}
\vspace{.0 cm}
\end{table*}

Table~\ref{table:lit} summarizes the experimental parameters of several studies from literature. Here we present a modified Scanning Stereoscopic Particle Image Velocimetry (SSPIV) that is suited for large depths of scan~(100 mm), with scanning speed reaching $10^4$ mm/s or higher. We used a thin, light rotating acrylic prism, which enables fast rotational speeds, up to 5000 rpm. As compared to prior works \citep{hoyer20053d,casey2013scanning,david2012time} where an expanded laser sheet is deflected, here, a laser beam is laterally shifted first, followed by its passing through a cylindrical lens with a length equal to the scanning distance. This yields a setup that is adaptable and scalable for scan-speed and timing changes, and at the same time fairly simple in construction. We used a quintic spline based interpolation scheme, which minimizes the error in 3D reconstruction. The  method is validated using two flow setups: 1) a round jet at Re = 1000, and 2) the flow around a rising spherical particle at Re$_p$ $\approx$ 260. Finally we discuss possible extensions of the technique and some limitations.

\section{Experimental apparatus}

\begin{figure*}[!h]
\centering
\includegraphics[width = \textwidth]{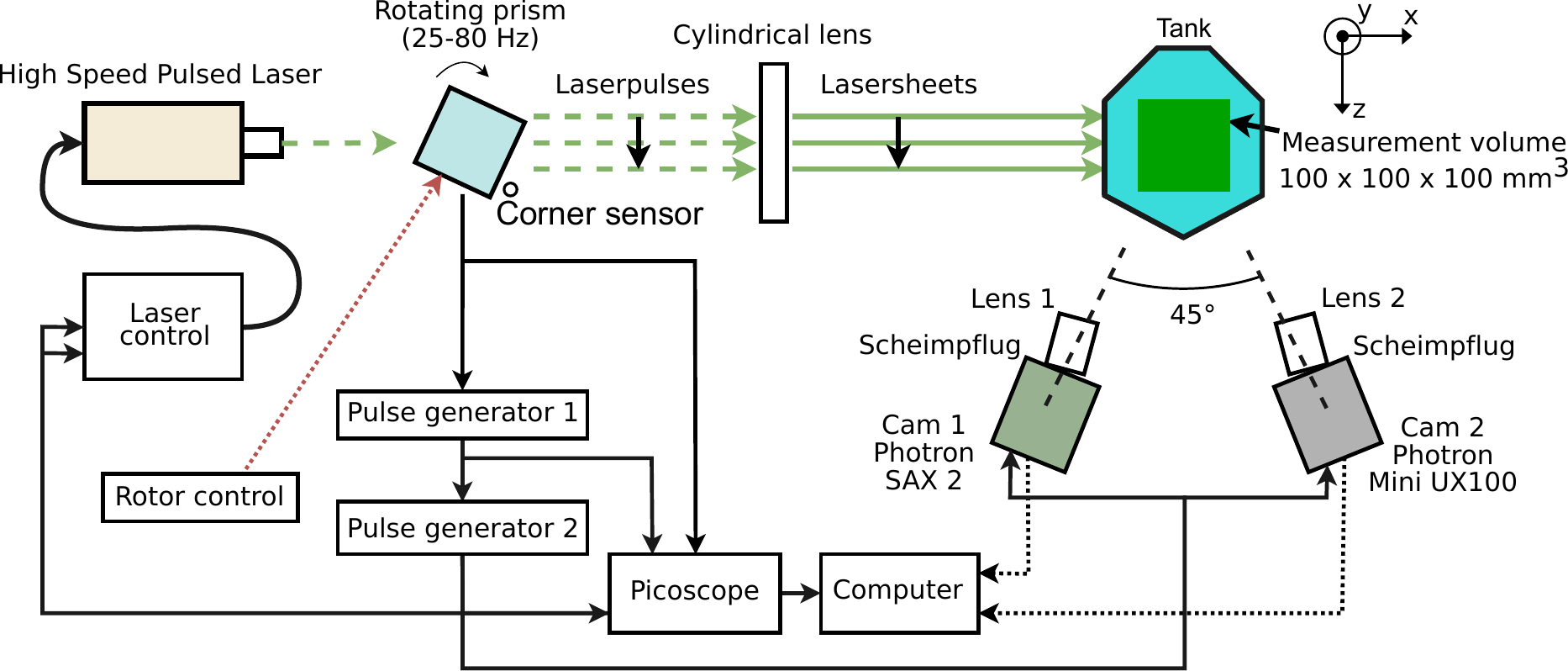}
\caption[Schematic overview of the Scanning Stereoscopic PIV setup.]{Schematic overview of the Scanning Stereoscopic Particle Image Velocimetry~(SSPIV) system.}
	\label{fig:setup_overview}
\end{figure*}

Figure~\ref{fig:setup_overview} shows a schematic of the SSPIV system. The experiments were conducted in an irregular heptagonal tank (width 400 mm, height 700 mm). The tank was designed to have adjacent faces at angles of 45$^\circ$, 60$^\circ$, and 90$^\circ$, of which the 45$^\circ$ angle was found  to be appropriate. The tank was positioned on a vertical stage that allowed movement in the y-direction. The tank was filled with  water or water glycerol mixture for refractive index matching, and seeded with 50 $\mu$m polyamide tracer particles~(Dantec PSP).

 The target measurement domain was a cubical volume of 10$^6$ mm$^3$ (see green shaded rectangle in figure~\ref{fig:setup_overview}).  Illumination was achieved using a High-Speed Pulsed Nd:YLF laser (527 nm, 30 mJ/pulse) with beam focusing optics. The maximum power was $100$ W, with pulse duration of $150$ ns.
The laser beam was deflected by refraction through the rotating acrylic prism ($200 \times 200 $ mm$^2$, and 5 mm height).  
Due to the difference in refractive index of air ($n_1 = 1$ and acrylic $n_2 = 1.49$), the beam gets refracted twice, causing it to be deflected laterally. The (relative) deflection of the laser beam by the prism is plotted against the prism angle in figure~\ref{fig:pulse_train}(a).
Using a cylindrical lens, the refracted beam was expanded into a sheet. The cylindrical lens was made out of  a round glass cylinder with a diameter around 10 mm and length 120 mm. The deflected beam after being expanded into the light sheet passes through the experimental section. The beam focusing optics was tuned to obtain a sheet thickness of $2$ mm, which is thicker than that used in most conventional 2D PIV setups~\cite{stanislas1997practicalaspects}, but necessary to capture out of plane particle displacements. The laser light scattered by the particles was imaged by two cameras for stereoscopic imaging.

The imaging system consisted of two high-speed cameras, namely a Photron SAX2 (Cam 1) with a recording rate of 12.5 kHz at 1 Mpx resolution, and a Photron Mini-UX (Cam 2) with a recording rate of 4 kHz at 1.3 Mpx resolution. The cameras were placed at an angle of 45$^\circ$ between each other.
Makro lenses with {100 mm (Cam 1) and 50 mm (Cam 2)} focal length were used, and the lens apertures were set to a minimum to ensure a large depth of field. Scheimpflug adapters were fitted to both lenses minimizing out-of-focus effects.

The rotational speed of the prism was controllable in the range 0-80 Hz (0-4800 rpm). In the present work, the rotation rate was set at 25 Hz, which results in a volume scanning rate of 100 Hz. The corresponding scanning speed was 10 m/s, which  exceeds those in prior works~\cite{casey2013scanning,david2012time,hoyer20053d}. The laser and cameras were triggered as a function of the orientation of the prism. Triggering was done using a corner sensor (see figure~\ref{fig:setup_overview}). This sensor consisted of a continuous wave laser and a photo diode. As the corner of the prism passed the corner sensor a 5 V output was generated, which was fed to the pulse generators. Two pulse generators were employed (type Agilent 3320A, Function/Arbitrary waveform generator 20 MHz). Pulse generator 1 created a so-called envelope pulse and pulse generator 2 created triggering pulses for the laser and the cameras. 
The signals from the corner sensor and the two pulse generators were monitored through a Picoscope $-$ a USB based oscilloscope. The data on the Picoscope was viewed on a computer screen.

	\subsubsection{Timing and triggering}

\begin{figure*}[!h]
	\centering
    \includegraphics[width = 1.00 \textwidth]{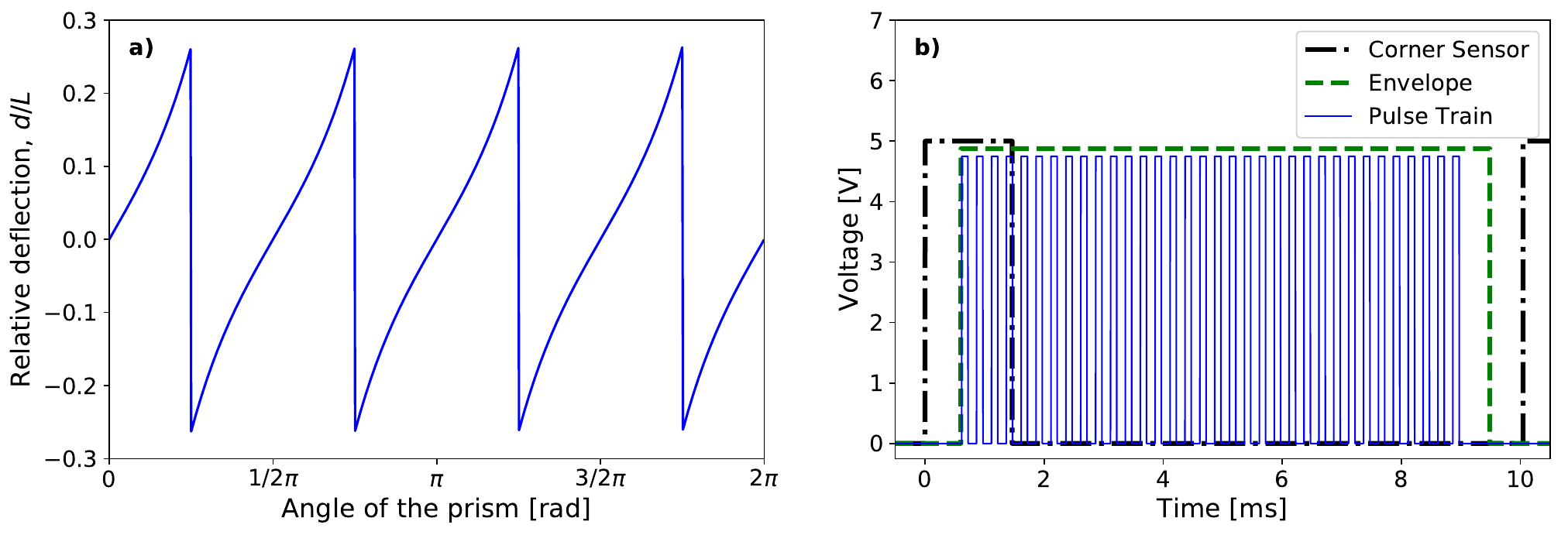}
	\caption{(a) Normalized deflection of the laser pulse as a function of the orientation of the prism. As a corner of the prism passes the laser pulse, the incident angle changes by $2\pi/n$ rad, for a n-sided prism. (b) Schematic representation of the signal from the corner sensor, pulse generator 1 (Envelope)  and pulse generator 2 (Pulse Train) are shown. First the corner sensor detects the prism, triggering pulse generator 1 to generate the envelope pulse. This pulse triggers pulse generator 2, generating the pulse train for the laser and camera~1.}
	\label{fig:pulse_train} 
\end{figure*}

The prism rotation rate was nearly constant, with a standard deviation of 0.1 Hz. The signal from the corner sensor triggered pulse generator 1, which then generated a very long pulse after a short delay $-$  the pulse being longer than the intended duration of the volume scan.  As the pulse is being generated, pulse generator 1 cannot be triggered again. This filtered out false starting triggers coming from the corner sensor as a result of the high powered laser scattering on the prism.  The long pulse from pulse generator 1 triggers pulse generator 2. The output from pulse generator 2 comprised of 34 successive pulses at a rate of 4 kHz. The signals were collected by the Picoscope, the resulting image is shown in figure~\ref{fig:pulse_train}(b).

The initial idea was to use a pulse generator to obtain two similar pulse trains; one for the two cameras and one for the laser. However, after analyzing the images of the first set of experiments, flickering was detected in the images. This decreases the quality of the processing. This was identified to be due to differences in capability of the internal trigger speed of the cameras. Hence, this method, although in principle simple, was not adopted.
The best settings for the Photron cameras (SAX2 and Mini-UX100) were as follows.
For camera 1~(Photron SAX2) the optimal setting was to use `{\it random reset trigger mode}'. In this case the timing of the shutter is reset, causing the pulses of the laser to temporally align better with the shutter. This simple adjustment removed the flickering issue of the SAX2 model. The same setting, however, was unsuited for camera 2~(Photron Mini UX100). In this case, the first pulse train was used to trigger a recording at 4000 fps, without feedback. Rare timing errors occurred, but were easily filtered out by thresholding the intensity profile of the recorded images. Processing of the intensity profiles will be detailed later.

\begin{figure}[!tbp]
	\centering
	\includegraphics[width = 0.5 \textwidth]{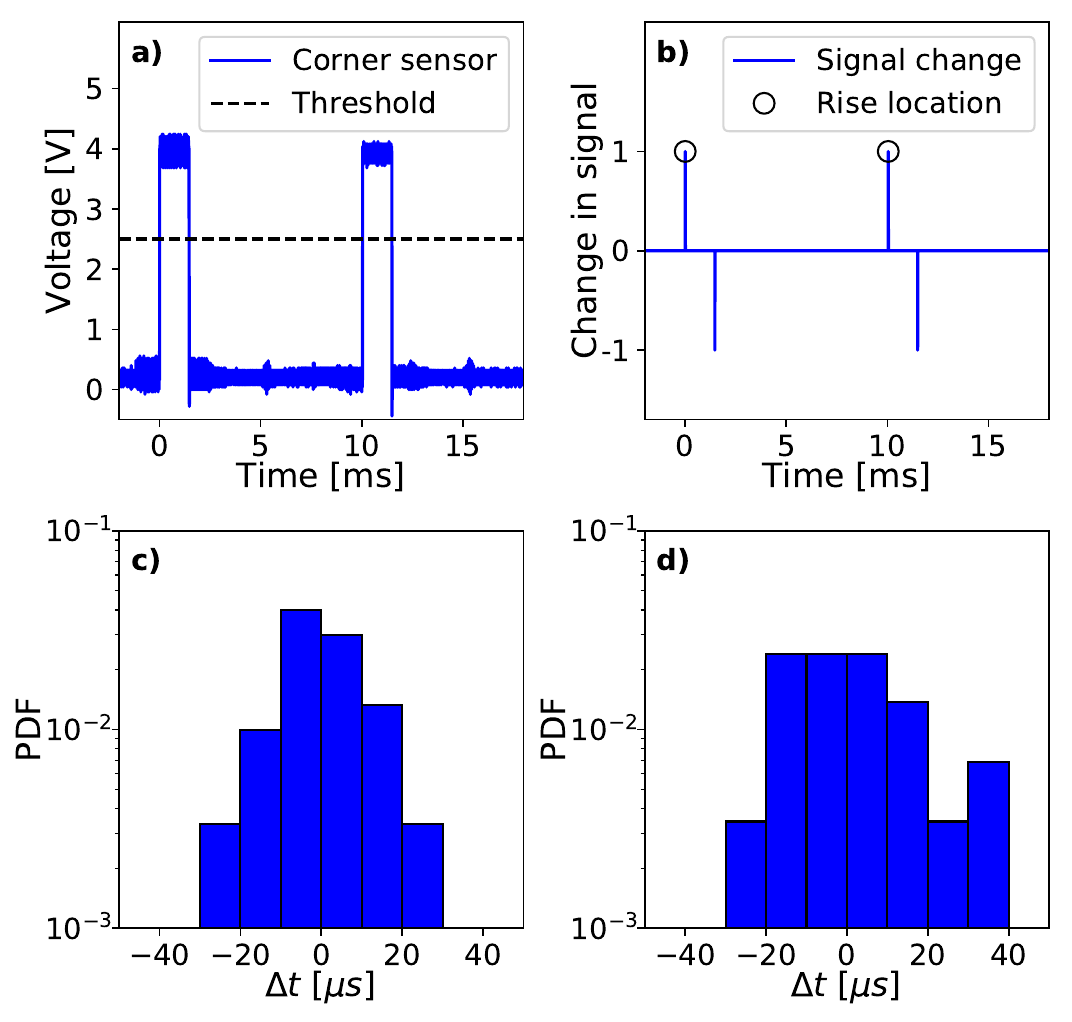}
	\caption[Processing the oscilloscope data to find the timing distributions.]{The distribution in timing between two adjacent detected corners. In (a) the voltage threshold that is used is shown. In (b) the change in the thresholded signal is shown, as the thresholded signal contains only $0$'s and $1$'s, the change is either $-1$, or $0$, or $1$. The value $1$ indicates a newly detected corner. This is used to calculate the period of the rotation. \textcolor{black}{In (c) the probability distribution of timing after subtracting the mean is given. (d) shows the difference between two successive periods, which results in a slight location difference between two successive (in time) sheets.}}
	\label{fig:prism_signal} 
\end{figure}

To obtain good correlations in the PIV recordings, the scanned laser sheet locations were required to reappear on overlapping planes. Since the timing is fixed, the stability of the location is solely a function of the stability of the rotation of the prism. The periodicity in the prism rotation is shown in figure~\ref{fig:prism_signal}. To this end, a reliable threshold in voltage was first set (figure~\ref{fig:prism_signal}(a)). The precise time of rise of the signal was then identified using the threshold signal (figure~\ref{fig:prism_signal}(b)), which was then used in the collection of periodicity data. The deviations in the timing of the corner sensor from the expected time interval are summarized as a probability distribution function in figure~\ref{fig:prism_signal}(c). Note that even a deviation of 30 $\mu$s ($\sim 10\%$ of volume scanning time) is highly improbable, with $P(30$~$\mu\text{s}) \approx 3 \times 10^{-3}$. A similar low probability of timing error is found in the successive time scans (figure~\ref{fig:prism_signal}(d)). Due to the non-linearity in the deflection, a difference in the timing does not give a fixed displacement error. The theoretical maximum deflection error arising from this timing difference is $0.45$ mm, assuming a rotational velocity of the prism of $25$ Hz. Additionally, we checked the sheets used in computing the cross-correlation. The same particles visible in the first image were (almost) always visible in the second image used for the PIV cross-correlation.
	
    \section{Calibration}

	The procedure followed for calibration of each image plane plays a crucial role in the quality of the results with SSPIV. A 3D calibration was performed by imaging a double-plane target plate incrementally
	moved in the scanning direction and later fitting the trajectory of each marker of the calibration target. Using the trajectories of the markers, a virtual calibration plate image was constructed at the exact location of each laser sheet.
    
\begin{figure*}[!tbp]
\centering
\includegraphics[width=\textwidth]{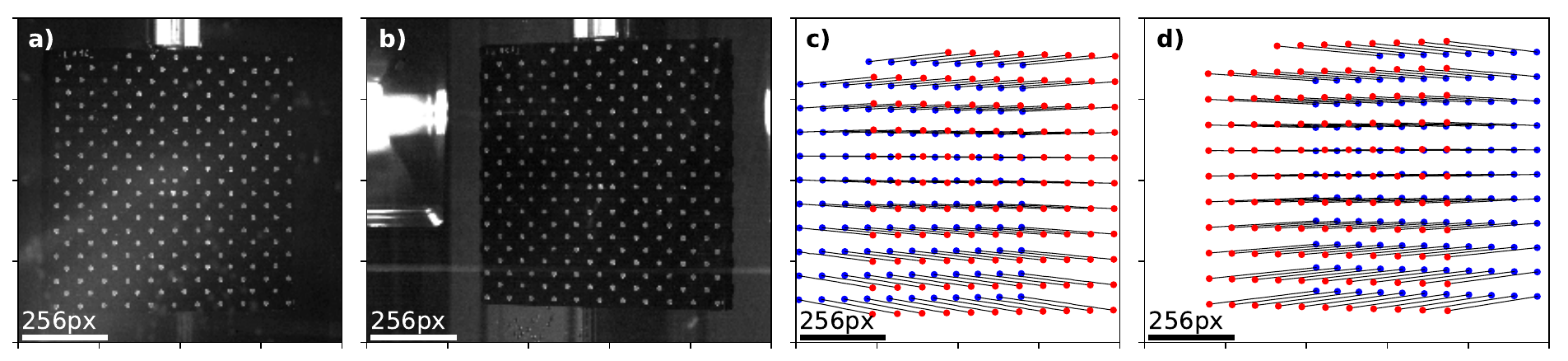}
\caption{(a)--(b) Images of the calibration plate (Lavision type~11), as viewed by the two cameras. (c)--(d) Trajectory of markers the traverse moves in the depth direction~(z coordinate) for (a) camera 1 and (b) camera 2. Each marker moves from its respective blue dot to the connected red dot as the calibration plate is moved towards the cameras.}
	\label{fig:calib_im}
\end{figure*}

    In order to perform the calibration over a large depth, the camera aperture had to be set at a low value. Representative images of the calibration target plate, as viewed by the two cameras, are shown in figure~\ref{fig:calib_im}(a) and~\ref{fig:calib_im}(b). The calibration plate was scanned at increments of $5$ mm over the whole range covered by the laser sheets. Using this method, $22$ images were collected for each camera, enabling a reconstruction of each marker's trajectory as a function of the z-location of the calibration plate. During the export of the calibration, a file containing all the pixel- and `real-world` positions is generated, including an index number based on the x-, y-, and z-positions.

\begin{figure*}[!tbp]
	\centering
	\includegraphics[width=\textwidth]{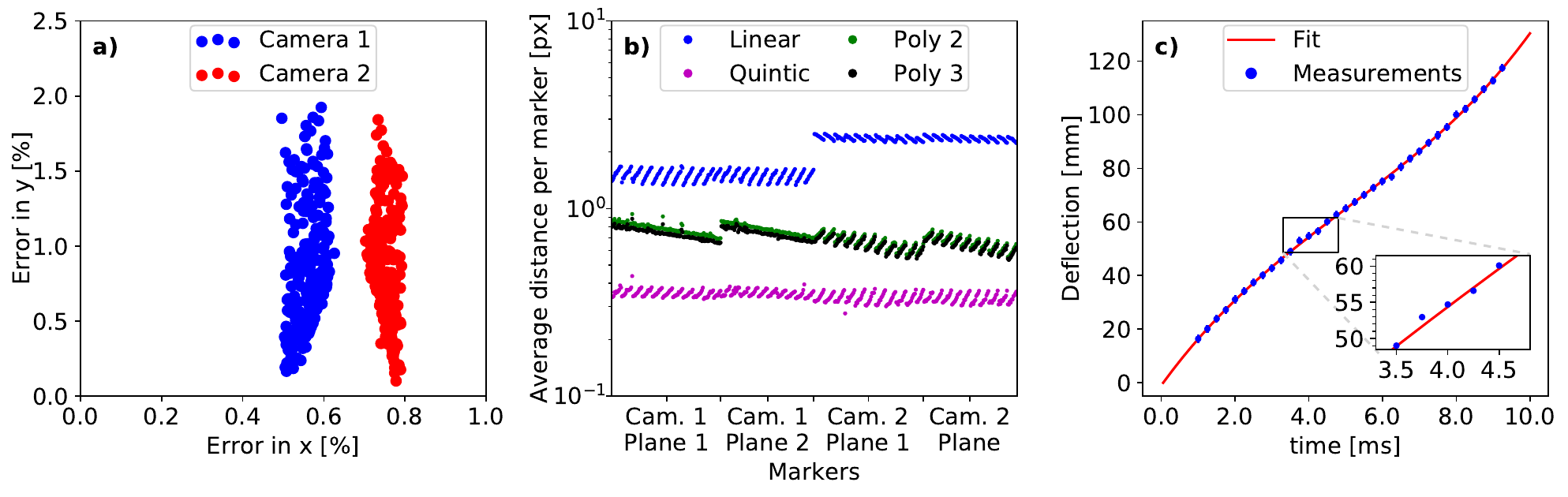}
	\caption{Average distance between fit and measurement for the movement of the markers on the calibration target. In (a) the result of a linear fit is shown. The average x and y difference for each marker is normalized by the total movement in x and y. (b) demonstrates the average distance for each calibration marker for different fitting methods. The calibration scale is roughly 10 px/mm. (c) The measured locations of the sheets are shown. To reduce the error in positioning, the curve is fitted with a model of the sheet deflection.}
	\label{fig:calib_trajectory_fits}
\end{figure*}

Using the parsed files, a reconstruction of the trajectories was made. An example of these trajectories is given in figure~\ref{fig:calib_im}(c) and~\ref{fig:calib_im}(d). The movement of the markers might appear to be almost linear in the figure; however, this is in fact not the case. The trajectory of each marker was fitted. First, a linear fit was tried, but this gives significant relative errors as shown in figure~\ref{fig:calib_trajectory_fits}(a). In \mbox{figure~\ref{fig:calib_trajectory_fits}(b)} the results obtained using various fitting methods are shown. Based on this result, a quintic fit was chosen, as the error between fit and marker points was sub-pixel. Note that this is not a true error but likely arises due to small errors in the prescribed locations of the calibration target.
For good interpolation, the precise locations of the sheets are needed. Locations of the sheets are determined by placing a piece of paper against the tank and marking them. To reduce errors in this step, a theoretical deflection profile was fitted through the measured locations, as shown in figure~\ref{fig:calib_trajectory_fits}(c). Inset to the figure shows a zoom-in view of the data vs. fit; the standard deviation in the fitting error was $0.4$ mm and the maximum error was $1.2$ mm.

\section{Reconstructing 3D-3C velocity field}   \label{calib} 

Certain features of the laser intensity were made use of in numbering the laser scans and performing the cross-correlation operations. Q-switching operation causes the laser intensity to have slight variations during the volume scan. The image intensity profile of each recording was analyzed by adding all the pixel values of each image. This resulted in an integer representation of the total intensity of the image. Repeating this for each image in a recording and scaling by the maximum image intensity gave an estimate of the intensity profile vs. time during the ``average'' volume scan. The variation in intensity showed a distinguishable pattern. The first laser pulse, generated after the corner sensor signal, was always the brightest. In subsequent laser pulses, the laser power was not perfectly maintained. These variations do not influence the quality of PIV, as the cross-correlation operation is only weakly affected by the relative intensity of the images pairs.
	
From the intensity profile, the first sheet of each volume illumination was found, and the time (amount of frames) between these `first' sheets was calculated. In some instances, the time between two `first sheets' was larger than expected, which indicated that the first sheet from a volume scan in between was not found in the intensity profile. This might have been due to issues with the internal timing of the shutter of camera 2 or the used trigger signal. When a missing `first' sheet was detected, to avoid mistakes in the post-processing, all images from the corresponding volume scan were rendered black. These completely dark images yielded obvious errors in the PIV processing. However, these could be conveniently filtered out and did not impact the overall quality of the time-resolved measurements.

The intensity profile was used to guide the sorting process, as this could pin-point the `first sheet' of each volume scan. For each sheet a PIV-project was created and in each project the two corresponding generated calibration images were processed. The sorted data from both cameras were imported into their respective projects and processed. Taking into account the basic design rules for PIV measurements \cite{springer_handbook}, an interrogation size of $24 $ px $ \times 24$ px was chosen, and all data in the project were processed. This gave two sets of processed and calibrated 2D-2 Component (2D-2C) recordings. The two 2D-2C recordings were then merged into a `multi-frame' and processed into one 2D-3 Component (2D-3C) recording. After repeating this process for all of the analyzed sheets, a 3D 3 Component (3D-3C) velocity field was generated. The data from all of the sheets were then combined into a volume of points and interpolated on a regular grid in order to estimate the spatial derivatives~\cite{prasad2000particle}. The values of 1030301 points were found using a barycentric interpolation of the simplex found by Delaunay triangulation. This meant that, first, the original volume was divided into tetrahedrons. Next, it was determined in which tetrahedron each interpolation point was located. The value was then determined by a weighted average based on the distances from the interpolation point to the vertices of the tetrahedron.

For each interpolation point, the indices of the four vertices of the tetrahedron were saved and used with the corresponding weights. This step, which takes an enormous amount of time, needs to be performed only once. The stored arrays are reused for every step of the recording. The initial phase takes approximately 1.5 hours, with each interpolation step thereafter completed in about 2 s. The total amount of data points that has been processed is 1030301 points per volume, with 213 volumes per recording, and 21 such sets, i.e. roughly $4~\times~10^9$~(4 billion) data points. The reduction in computational time compared to repeatedly calling a standard interpolation function was about 1.5 hours per 3D volume. For one recording with 213 volumes, the resulting total reduction in time was 197; in other words, a reduction from 9 days to about 1.6 hours. Next, the velocity components, $\vec{u}=(u,v,w)$, were directly obtained from the 3D reconstruction. Here, $u$, $v$, $w$ are the velocities along the x, y, z directions, respectively. Following this step, all 9 spatial velocity gradients were computed. A least-squares-based minimization was performed on the spatial derivatives, with net divergence minimized.

In the following sections, two experiments performed with the 3D-SSPIV system will be described. Section \ref{roundjet} will discuss the flow field due to a round jet\cite{pope2001turbulent} at a jet Reynolds number Re = 1000, and Section \ref{rise_sphere} will discuss the flow field around a freely rising sphere at a sphere Reynolds number, Re$_p$ $\approx$ 260.

    \section{Experiment I: Round jet} \label{roundjet}
\begin{figure}[!h]
    		\centering
    	\includegraphics[width= 0.85\columnwidth]{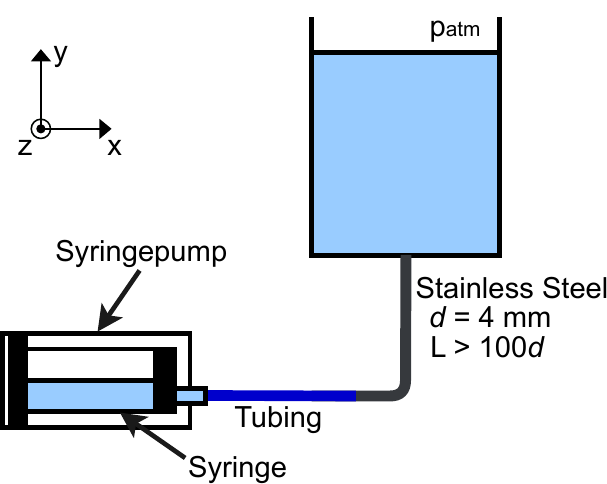}
		\caption[A simplified schematic of the round jet experiment.]{A simplified schematic of the round jet experiment. Water seeded with tracer particles is injected at a constant volume rate into the tank.}
		\label{fig:jet_setup_overview}
	\end{figure}

    The first experiment performed was a round jet flow at a jet Reynolds number of 1000.  Figure~\ref{fig:jet_setup_overview} shows a schematic of the experiment in a container filled with water. The setup comprised a glass tank of height $\approx$ 700 mm, and width $\approx$ 400 mm. The tank was filled with water, and a small circular orifice of diameter 4 mm was located at its base. The orifice was connected through a straight steel tube of 400 mm length, which was the developing length required for the issuing jet to not deviate laterally. Beyond this length, the tube was bent 90$^\circ$ and connected to a flexible tubing and a syringe pump. The syringe pump provided a controlled flow rate, which created a nearly vertical jet of the same fluid. The flow rate was controlled to result in a jet Reynolds number Re $\equiv u_m d/\nu = 1000$. Here $u_m = Q/(\pi d^2/4)$ is the mean bulk velocity, where $Q$ is the volume flow rate, $d$ is the orifice diameter, and $\nu$ is the kinematic viscosity of the fluid. During the first 15 s, roughly $1000$ dimensionless times based on $\tilde{t}= t/{t_{c}} = t/(d/u_{m})$ of injection by the syringe pump, recordings were not made in order to minimize the transient phase. Following this waiting period, SSPIV recordings were made.

     \begin{figure}[!h]
    	\centering
    \includegraphics[width = 0.5 \textwidth]{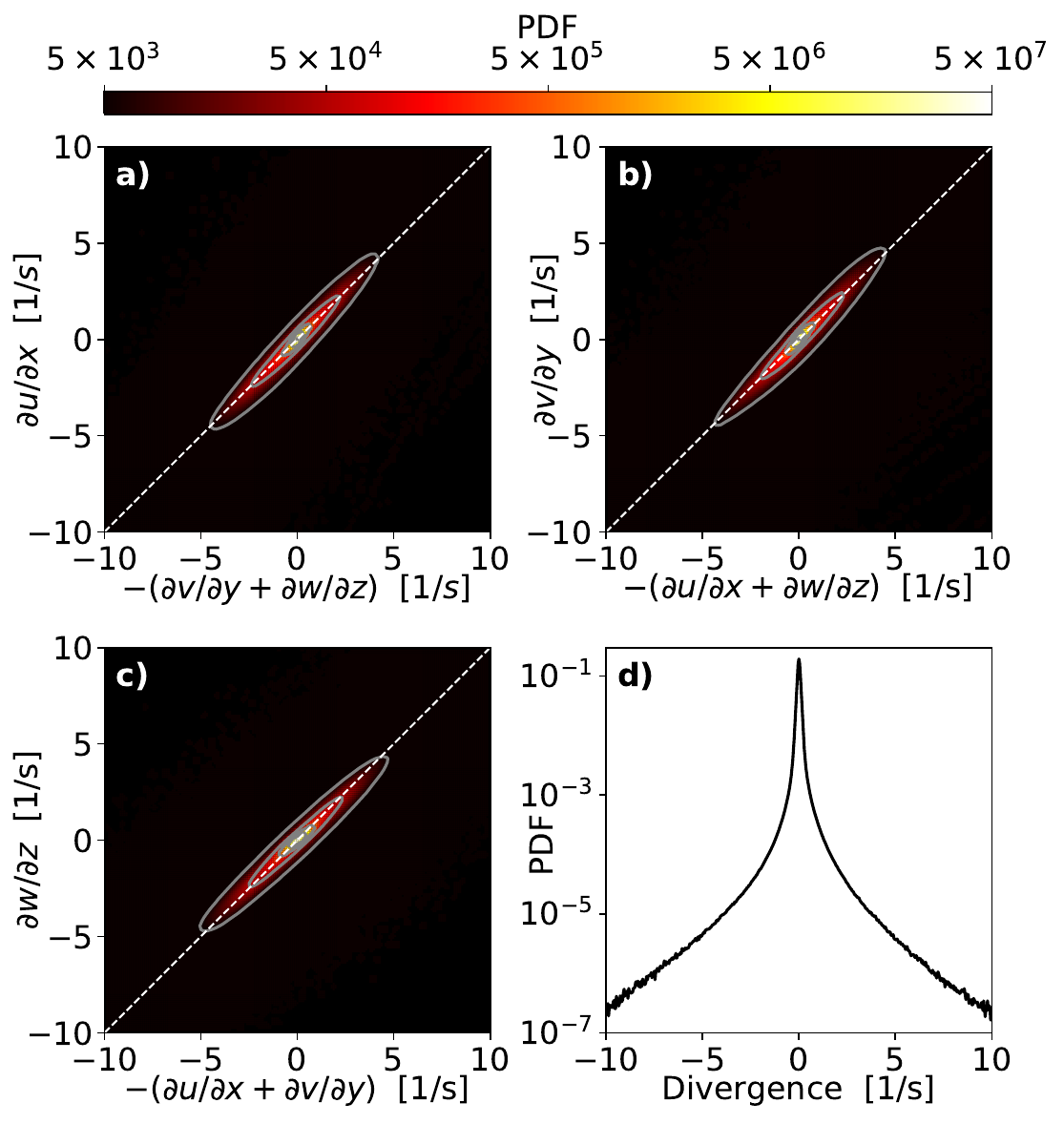}
    	\caption[Joint PDFs of divergence for a jet Re = 1000.]{Accuracy checks using joint PDFs of the divergence. (a)--(c) Joint PDFs of the one of the terms in the divergence $\nabla \cdot{\bf u}$ vs the negative of the sum of the two other terms. For incompressible flow~(divergence-free), the data should fall on the dashed line. Deviations from the dashed line with slope 1 suggest errors in the spatial gradients. (d) PDF of the divergence on semi-log scale.}
            	\label{fig:divergence_1000}
    \end{figure}
    
 The accuracy of the spatial velocity gradients can be assessed based on the divergence criterion: $\nabla \cdot{\bf u} = 0$, where ${\bf u}({\bf x},t)$ represents the velocity vector at position vector ${\bf x}$ and time $t$. Since the flow can be regarded incompressible,
 
 \begin{equation}
 \nabla \cdot {\bf u} = \frac{\partial{u}}{\partial{x}} + \frac{\partial{v}}{\partial{y}} + \frac{\partial{w}}{\partial{z}} = 0
    \end{equation}

In figure~\ref{fig:divergence_1000}, the joint probability density functions (PDFs) of the divergence terms are shown. The velocity gradient $\frac{\partial{u}}{\partial{x}}$ against $- (\frac{\partial{v}}{\partial{y}} + \frac{\partial{w}}{\partial{z}})$ is plotted in figure~\ref{fig:divergence_1000}(a), where the two terms constitute the terms contributing to $\nabla \cdot{\bf u}$. Points that are away from the dashed line correspond to non-zero divergence, and hence indicate measurement error. The velocity vectors obtained from Scanning Stereoscopic PIV follow the diagonal line fairly well. The contours are colored based on a log-scale, which meant that the outer-envelope (in white) of the PDF was nearly 4 orders less probable than the innermost contour.   Figure~\ref{fig:divergence_1000}(b--c) shows similar plots, but with the remaining two terms of $\nabla \cdot{\bf u}$ on the vertical axis. Here again, the deviations from the dashed line are low in probability. In figure~\ref{fig:divergence_1000}(d), the PDF of $\nabla \cdot{\bf u}$ is presented on a semi-log scale. Most of the data are distributed around $\nabla \cdot{\bf u} = 0$. Note that a divergence value of 1 [s$^{-1}$] ($\approx 1.33$ times the rms value) has a probability 1/$696$ of the peak value, which is comparable to other 3D PIV studies~\cite{elsinga2006tomographic,jodai2016experimental,casey2013scanning}.

\begin{figure}[!h]
   	\centering
   \includegraphics[width=0.5 \textwidth]{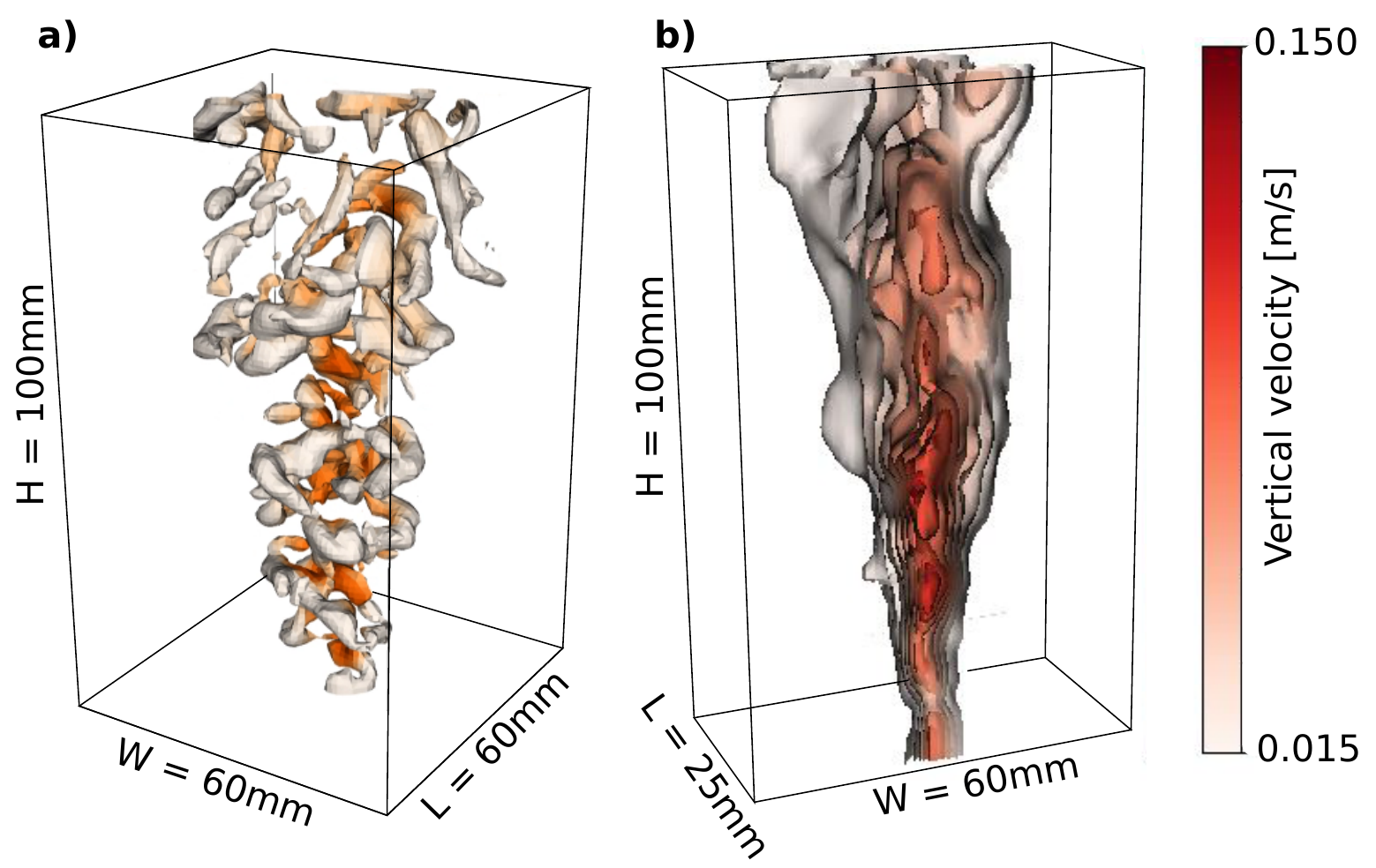}
    \caption[Snapshots of the instantaneous vertical velocity for a jet Re = 1000.] {Snapshots of the instantaneous vertical velocity of the Re = 1000 jet. From left to right the time increases by 0.2 s. The jet fluctuates in time. The time interval between successive snapshots is around $10 T_c$, where $T_c = d/v_\text{jet}$ is a characteristic timescale based on the orifice diameter $d$ and bulk velocity $v_\text{jet}$ at the nozzle exit.} \label{jet_instantaneous_1000}
\end{figure}

\begin{figure}[!h]
	\centering
\includegraphics[width=0.5 \textwidth]{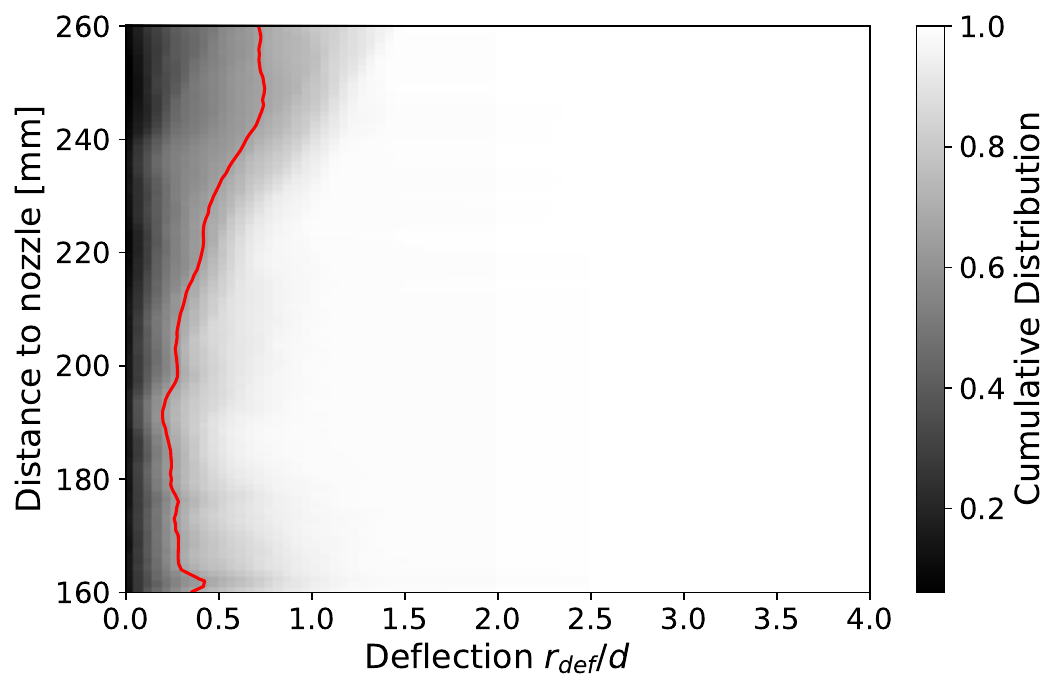}
\caption[Centerline deflection for a jet Re = 1000.]{Cumulative distribution of the deflection of the centerline of the jet, Re = 1000. The drawn contour line is at one standard deviation.} \label{centerline_deviation}
\end{figure}

Next, we evaluate features of the instantaneous flow field obtained from the SSPIV. Figure~\ref{jet_instantaneous_1000}a shows a snapshot of the instantaneous flow structures of the vertically issuing jet, based on the $\mathcal{Q}$ criterion. A supplementary video shows the evolving flow structures from the high-speed SSPIV. A cross-section of the jet is shown in figure~\ref{jet_instantaneous_1000}b, colored by the vertical velocity. The diverging angle of the jet is visible in this snapshot. While the instantaneous structures are highly unsteady and show lateral undulations in time, the mean jet is nearly vertical. We explore statistics of the horizontal deflection of the jet in figure~\ref{centerline_deviation}. The cumulative distribution of the deflection of the center-line of the jet is shown at various distances from the jet-origin. The standard deviation of the jet deflection, conditioned on the distance from the orifice, is given by the red curve. This lies in the range $[0.5d, 1d]$, suggesting that the jet remains stable during the full duration of the experiment. The standard deviation of the jet's axis fluctuation corresponds to an angular deviation of 0.6$^\circ$ from the center-line.

   \begin{figure}[!b]
    	\centering
    \includegraphics[width=0.5 \textwidth]{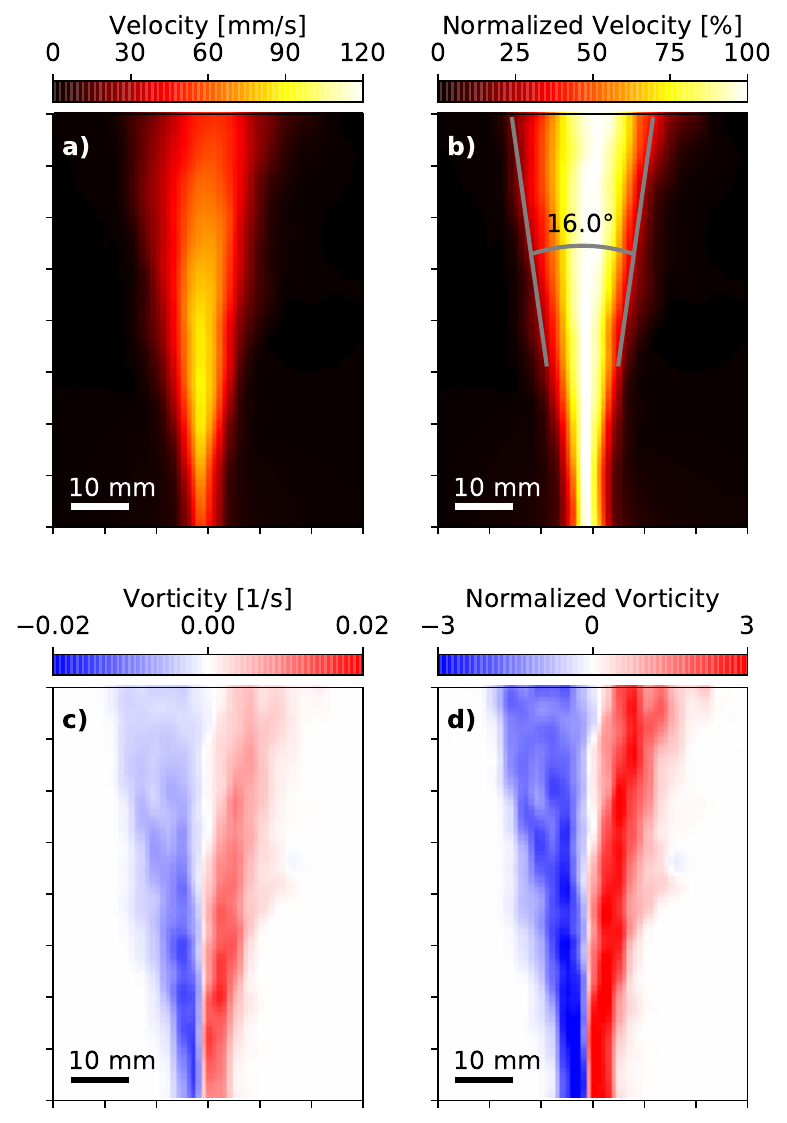}
    	\caption[Velocity and vorticity for a jet Re = 1000.]{Averaged velocity and the vorticity fields for the round jet at $\text{Re} = 1000$. (a) velocity of the jet, (b) velocity normalized by the centerline velocity, (c) out-of-plane vorticity, and (d) out-of-plane vorticity normalized by the rms value of the vorticity at each height. Normalizing as done in (b) and (d) reveals the nearly conical shape of the spreading of the jet.}
        \label{fig:averaged_jet_1000}
        \end{figure}

    \begin{figure}[!h]
    		\centering
    	\includegraphics[width=0.5 \textwidth]{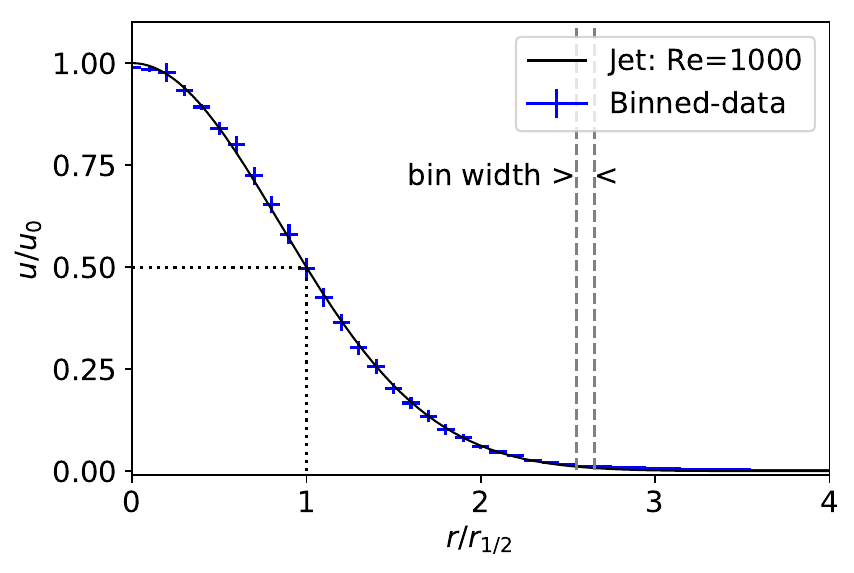}
    	\caption[Self-similarity match for a jet Re = 1000.]{Decay of normalized vertical velocity as a function of normalized radial position for Re = 1000 jet. The values are normalized by the jet centerline velocity and the jet half-width, which seems to collapse the curves at different heights. The data is compared to the self-similar profile of a developed turbulent jet.}\label{jetprofile_1000}
    \end{figure}

The average velocity and vorticity fields at the mid-plane of the Re = 1000 jet are displayed in figure~\ref{fig:averaged_jet_1000}(a--b) and (c--d), respectively. The field of view is approximately $15 d \times 20 d$, and the lower line is at a distance $40 d$ from the nozzle exit. The jet velocity magnitude (figure~\ref{fig:averaged_jet_1000}(a)) decreases with distance from the nozzle. However, one cannot appreciate a self-similar distribution here, as the centerline velocity also decreases with height. In figure~\ref{fig:averaged_jet_1000}(b), the velocity magnitude normalized by the centerline velocity is plotted. One can notice a nearly constant spreading angle, in agreement with prior works~\cite{labus1972experimental,pope2001turbulent,nguyen2024analysis}. However, the half-angle of spreading ($\approx$ 8$^\circ$) is lower than the 11.8$^\circ$ expected of a turbulent jet~\cite{hussain1994}. This is likely a finite Reynolds number effect. The average vorticity $\omega$ is zero along the centerline due to the inflection in the velocity profile (figure~\ref{fig:averaged_jet_1000}(c)). At larger distances from the nozzle exit, the vorticity decreases. Moving radially outwards, $\omega$ peaks and then decreases, yielding a double-peaked profile. The spreading of the vorticity field can be better appreciated when normalized by the standard deviation of $\omega$ along the height, as shown in figure~\ref{fig:averaged_jet_1000}(d).

As explained in the above section, the averaged flow field suggests a constant jet spreading angle in the upper half of the measurement section. Therefore, the velocity profile of the jet is examined in this region (shown in figure~\ref{jetprofile_1000}) where the jet velocity normalized by the centerline velocity is compared with the velocity profile for a turbulent jet~\cite{pope2001turbulent}. For a turbulent jet, sufficiently far away from the origin, the velocity of the jet should follow a law of similarity at all cross-sections, except for a stretching factor.  The data points match fairly well even though the Re in our experiment is not very high.

 \section{Experiment II: Buoyant sphere} \label{rise_sphere}
\begin{figure}[!h]
    		\centering
    	\includegraphics[width= 0.99 \columnwidth]{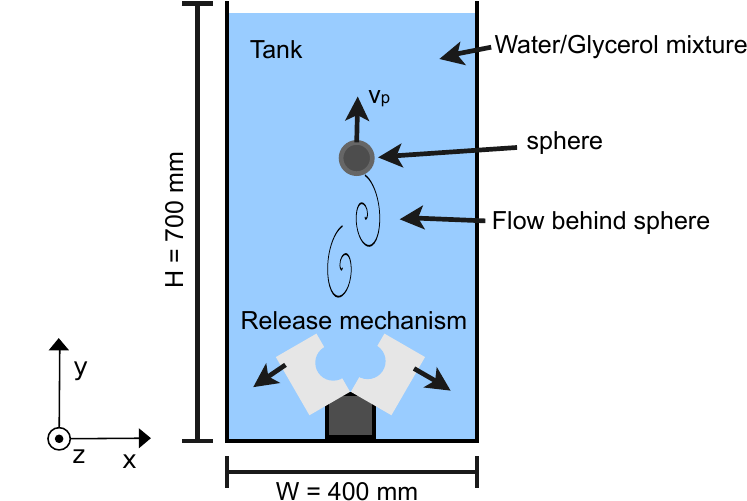}
	\caption[A simplified schematic of the rising sphere experiment.]{A simplified schematic of the rising sphere experiment. A pneumatically controlled release mechanism opens very gently to let the sphere rise without creating much disturbance to the flow. Experimental conditions: Sphere to fluid density ratio $\rho_p/\rho_f \approx 0.98$, and sphere Reynolds number Re$_p \approx$ 260. Note that $u$, $v$, and $w$ are the velocity vectors obtained along the x, y, and z directions, respectively.} \label{sphere_release_expt}
\end{figure}

Next, we discuss the second test experiment, which involves the flow around a marginally buoyant ($\rho_p/\rho_f \approx 0.98$) sphere rising  steadily at a mean sphere Reynolds number Re$_p\approx$ 260. Figure \ref{sphere_release_expt} shows a schematic of the experimental apparatus. The setup used was the same as in experiment 1: a glass tank of height = 700 mm and width = 400 mm. However, in this case, the tank was filled with a mixture of glycerol and water~(63\% volume fraction of glycerol, $\rho = 1175$ kg/m$^3$, $\nu = 1.68 \times 10^{-5}$ m$^2$/s). Fixed at the bottom of the tank was a pneumatically controlled mechanism for the sphere release. The mechanism could be opened very gently, thus providing precise control over the exact release time, and leaving very little disturbance to the fluid during the release process. 

 In order to illuminate the whole volume around the sphere without any shadows, it was important that the sphere remained transparent with a refractive index comparable to that of the surrounding fluid. Hence, hydrogel spheres were used for the experiment. By changing the duration of submersion in liquid, it was possible to tune both its density and refractive index. For refractive index matching, the sphere was immersed in a small tank of water-glycerol mixture, with a 4~mm square grid placed behind the tank. A 63\% volume fraction of glycerol in water mixture was chosen, which resulted in minimal distortions of the grid. 

The carrier fluid was seeded with $50~\mu$m Polyamide Seeding Particles with density matched to that of the glycerol-water mixture. As in Experiment I, here also, the experiment was designed to make recordings only after the initial transients during release were completed. To ensure this, the rising sphere was allowed to reach a velocity close to its terminal rise velocity in quiescent liquid. Hence, the chosen measurement volume for the PIV measurement was at a certain distance above the release location, in order for the sphere to have reached a quasi-steady state of rise.

We performed a rough estimate of the distance (height) required to reach 97\% of the terminal rise velocity. The dominant forces acting on the rising sphere were gravity, buoyancy, and drag. Therefore, the net force can be written as
\begin{equation}
F_{p_{sum}} = g \mathcal{V}(\rho_{f} -\rho_{p}) - \dfrac{1}{2}\rho_{f}v_{p}^2C_d A_{p}, \label{force_bal}
\end{equation}
where $g$ is the acceleration due to gravity, $\mathcal{V}$ is the sphere volume, $\rho_p$ and $\rho_f$ are the sphere and fluid densities, respectively,  $C_d$ is the drag coefficient, $v_p$ is the sphere velocity at any instant during the rise, and $A_p = \pi R^2$ is the projected area of the sphere, with $R$ being the sphere radius. Note that added mass and history force effects \cite{mathai2016microbubbles,mathai2018flutter,ern2012wake} were ignored in this analysis, and a constant drag coefficient based on the terminal Reynolds number of the sphere was assumed. Equating the net force to be zero gives an expression for the terminal rise velocity
\begin{equation}
v_{\text{p}_\text{term}} = \sqrt{\dfrac{8g R}{3C_d}\dfrac{(\rho_{f}-\rho_{p})}{\rho_{f}}}\\ \label{vterminal}
\end{equation}

From Eq.~\ref{force_bal}, the acceleration of the sphere may be written as:
\begin{equation}
\label{eq:a8}
\dfrac{dv_{p}}{dt} = \dfrac{g \mathcal{V}(\rho_{f} -\rho_{p}) - \dfrac{1}{2}\rho_{f}v_{p}^2C_d A_{p}}{\rho_{p} \mathcal{V}} 
\end{equation}
The above equation can be expressed in the form $\dot x = a + b x^2$, for which a standard solution exists, assuming \(x_{t=0} = 0\), given by 
$x(t) = \sqrt{\dfrac{a}{b}}\tan(t\sqrt{ab})$. This results in the expression for the rise velocity
\begin{equation}
\begin{aligned}
v_{p}(t)=& v_{\text{p}_\text{term}}\tanh{\bigg(\dfrac{t}{t_0}\bigg)},
\end{aligned}\label{vpt}
\end{equation}
where $t_0^{-1} = \sqrt{\dfrac{3C_d}{8R}g \frac{\rho_{f}^2}{\rho_p^2} \dfrac{(\rho_{f}-\rho_{p})}{\rho_{f}}}$ is the time constant. From this estimate of the velocity evolution, the distance traveled by the sphere in time $t$ can be written as $H_p = \int_{0}^{t}v_{p}(t)dt$. By invoking the relation $\int \tanh(c_1x)dx=\dfrac{1}{c_1}\ln(\cosh(c_1x))$, Eq.~\ref{vpt} can be integrated analytically, which yields the distance traveled as

\begin{equation}
\begin{aligned}
H_\text{char} =& \dfrac{8R}{3C_d}\dfrac{\rho_{p}}{\rho_{f}}\ln\bigg(\cosh\big(\dfrac{t}{t_0}\big)\bigg).\\
\end{aligned} \label{distravel}
\end{equation} 

From Eq.~\ref{vpt}, one can obtain the time to reach 97\% of the terminal velocity, $\dfrac{t_{97\%}}{t_0} = \tanh ^{-1}\big(0.97\big)$. Substituting this value into Eq.~\ref{distravel}, we obtain a characteristic distance $H_\text{char} \approx 100$ mm. We have used the sphere mass density $\rho_p/\rho_f = 0.98$, sphere radius $R = 22$~mm measured in our experiment, and assumed a mean drag coefficient $C_d = 0.75$ based on the terminal Reynolds number range in the experiment. The measurement volume of the SSPIV was chosen to be just above $H_\text{char}$.

\begin{figure}[!h]
 	\centering
 \includegraphics[width = 0.5 \textwidth]{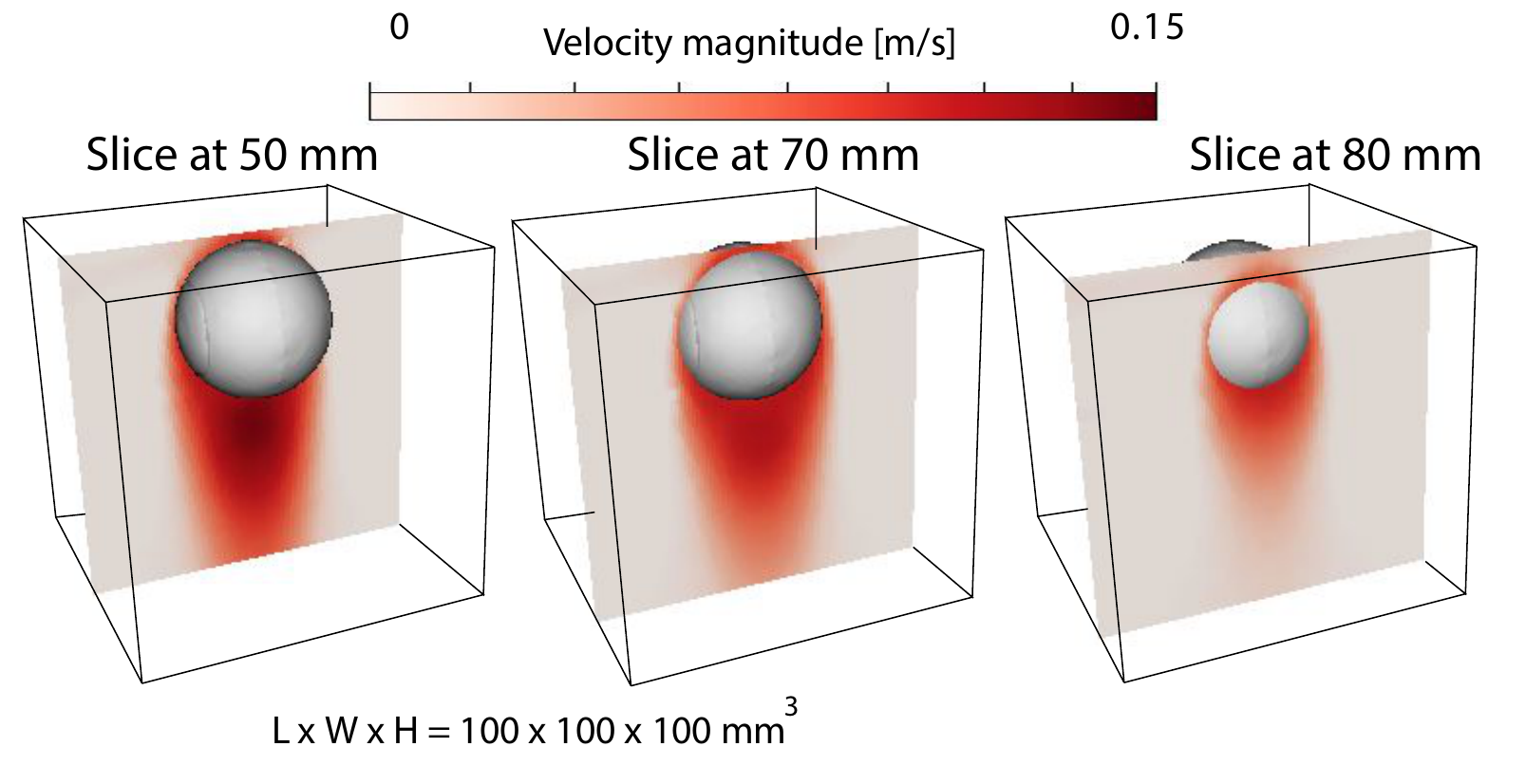}
    \caption[Velocity magnitude around the rising sphere.]{Velocity magnitude around the rising sphere, at different z locations slicing the sphere. Here the slide location in mm corresponds to the distance from the boundary of the PIV domain as measured from the opposite of the viewing side. Flow quantities (velocities and vorticity) in the reference frame of the tank.} \label{vslices_sphere}
\end{figure}

The grid points overlapping with the sphere volume were removed from the PIV analysis. To this end, the trajectory of the sphere was first tracked by using the 2D slices at  different locations in the volume, and using circle detection to locate the center and surface of the sphere. By using three orthogonal slices, the exact (x,y,z)-location of the sphere was found. This procedure was repeated for the entire time series of volumes to obtain a time-varying mask for the PIV data analysis.

\begin{figure*}[!h]
	\centering
\includegraphics[width=1.0 \textwidth]{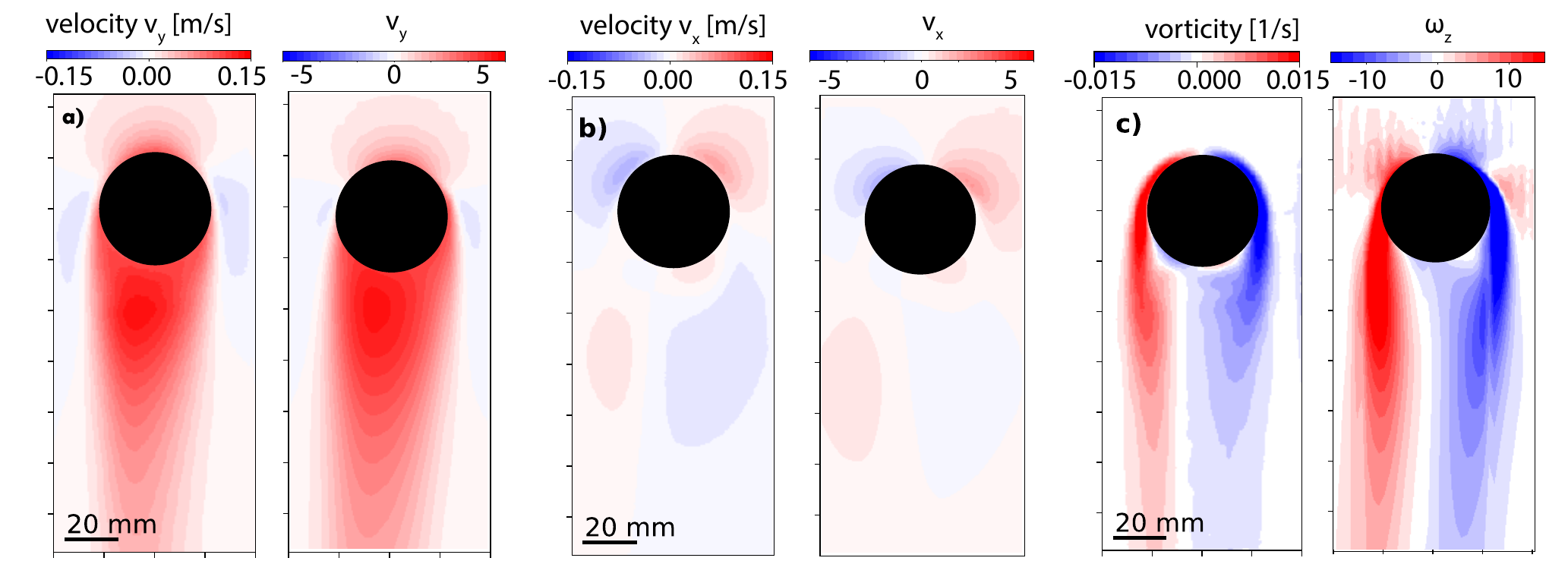}
    \caption[Velocity field and streamlines around the sphere.]{{\color{black}Experimental {\it(left)} and simulated  fields {\it(right)} of rising sphere and settling droplet, respectively, at comparable Reynolds number (Re$_p$ = 260 \& Re$_d$ = 262, resp.) are shown in each panel: (a) vertical velocity; (b) horizontal velocity, (c) out-of-plane vorticity.}}\label{fig:sphere_flow_spherecoord}
\end{figure*}

Figure~\ref{vslices_sphere}(a) shows vertical planes at different z-locations slicing through the sphere. The slices are colored by the velocity magnitude. The wake behind the sphere is clearly visible, the extent of which is farthest for the mid-slice through the sphere (left figure), and is reduced for the other two slices~(center and right figures). The size of the wake region is largest in the mid-slice. Moving downstream, the wake-disturbance decreases and it has decayed significantly by about 120 mm below the sphere, which translates to about 2.7 sphere diameters.

We now focus on the trajectory of the rising sphere. Interestingly, the sphere follows an oblique path at an angle of 4$^\circ$ with the vertical. Despite the inclined trajectory, the sphere wake flow is surprisingly steady. From literature on flow past a sphere {\color{black}at Re$_p \approx 250$}, we note that the wake is steady,  but no longer axisymmetric \cite{johnson1999flow}. However, a plane that slices along the inclined line of the sphere trajectory can be constructed, on either side of which the flow will have mirror symmetry. On this interpolated plane, the flow field conditioned on the instantaneous position of the sphere is performed. 
\begin{figure}[!h]
	\centering
\includegraphics[width=0.4 \textwidth]{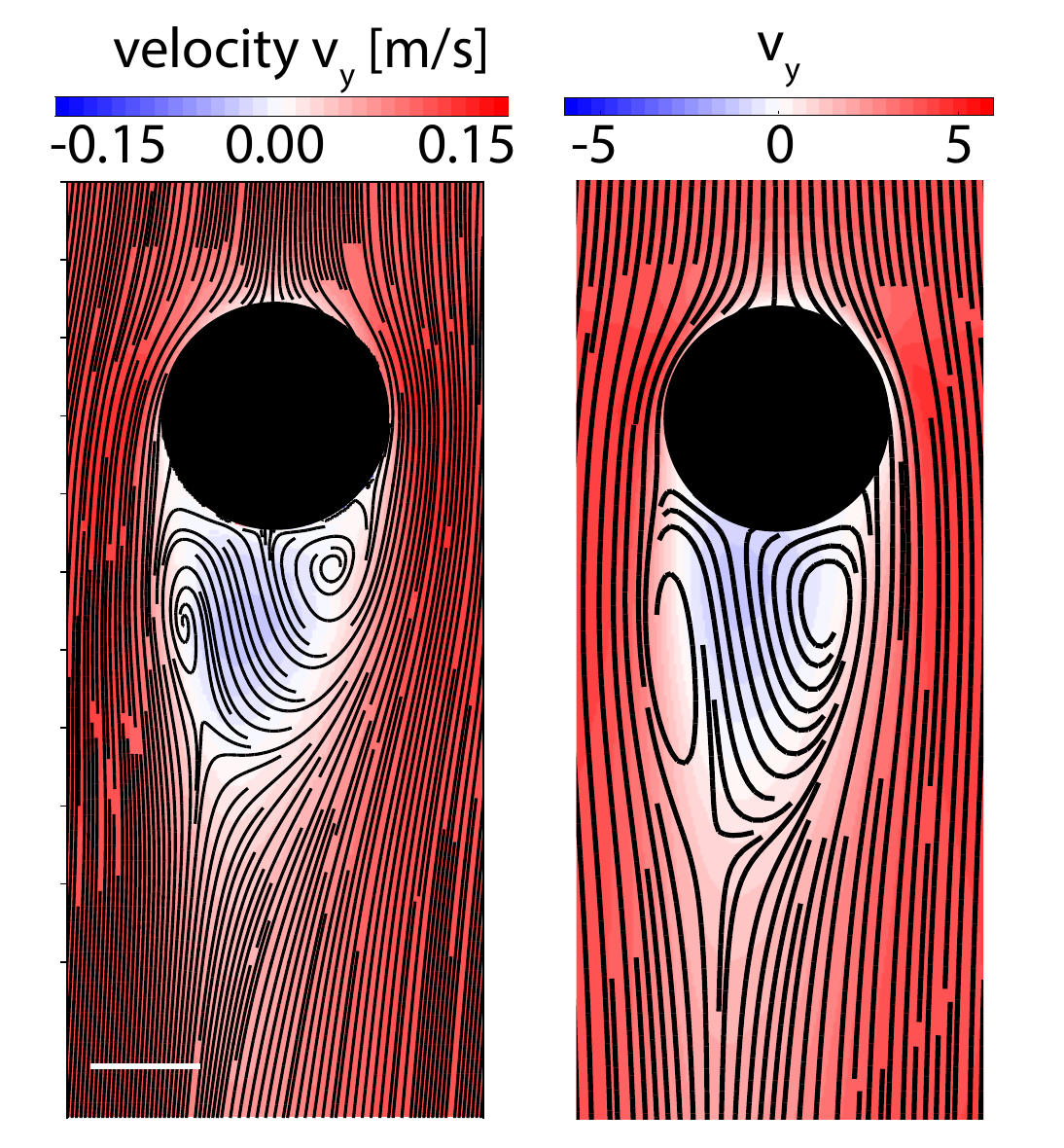}
    \caption[Velocity field and streamlines around the sphere.]{\color{black}Experimental (left) and simulated(right) fields of vertical velocity with streamlines overlaid. Here, the particle velocity was subtracted to help visualize the wake structure.}\label{fig:streamlines}
\end{figure}
\subsection*{Comparison with numerical simulations}
{\color{black} Direct numerical simulations of a droplet sinking in quiescent fluid are also made to compare with experimental results. Since the wake is not unstable at the studied Re, we expect the flow fields to be somewhat comparable for the rising sphere and the settling droplet. Additionally, we wish to highlight any subtle differences due to density difference at fixed Reynolds number. The simulation has terminal Reynolds Re$_d$ = 262, which is close to the experimental Reynolds number~(Re$_p\approx$~260). {\color{black}The simulation was  performed with a tree-based adaptive solver for the incompressible Navier-Stokes equations in complex geometries similar to Gerris; the solver can be found as `navier-stokes/center' in} Basilisk, an open source hydrodynamic simulation tool \cite{popinet2003gerris,popinet2024basilisk}. In the simulation, a droplet whose density and viscosity was 10 times and 100 times of surrounding fluid respectively {\color{black}was released}; the droplet, after release, accelerated until it reached a quasi-steady state where Re$_d$ = 262. In order to compare with the experimental results, the figures of simulations were flipped upside down since the droplet in the simulation was sinking while the sphere was rising}.
{\color{black}In figure~\ref{fig:sphere_flow_spherecoord}; the left figures of each panel are the experimental results:}
{\color{black}Figure~\ref{fig:sphere_flow_spherecoord}(a)}{\it(left)} shows the vertical velocity field in the lab-coordinate-system. The {\color{black}moving} sphere drags the surrounding fluid with it. The asymmetry of the wake becomes obvious once the horizontal velocity field is plotted, as shown in {\color{black}figure~\ref{fig:sphere_flow_spherecoord}(b){\color{black}{\it(left)}}}. While the flow just above the sphere shows left-right symmetry, the wake below the sphere has a clear breaking of symmetry. The same can be said about the {\color{black}vorticity field (figure~\ref{fig:sphere_flow_spherecoord}(c){\color{black}{\it(left)}})}. The region of negative vorticity (blue) is noticeably larger than that of the positive vorticity (red). This symmetry breaking of the horizontal velocity and vorticity fields is consistent with the observations of an oblique trajectory of the sphere.

{\color{black}The right-panels of the sub-figures~\ref{fig:sphere_flow_spherecoord}(a)-(c) are the DNS results for the droplet. All simulation figures were normalized by the maximal value of the field instead of using real physical world units.}
{\color{black}  {\color{black}
In figure.~\ref{fig:sphere_flow_spherecoord}(a){\color{black}{\it(right)}} and (b){\color{black}{\it(right)}}; vertical and horizontal velocities were plotted and compared. Non-axisymmetry of the wake flow was observed. There are minor differences between experiments and simulations, especially in the horizontal velocity slice (Fig.~\ref{fig:sphere_flow_spherecoord}b{\color{black}{\it(right)}}). This could be caused by the difference in the density ratio of the droplet and the sphere. The maximal difference in field value at any point of two cases is about 7\%, which is relatively small, and we can conclude that the simulated flows are remarkably similar to the experimental fields. In figure.~\ref{fig:sphere_flow_spherecoord}(c){\color{black}{\it(right)}} the results of simulation of vorticity have been shown; the vorticity fields were comparable {\color{black} to experimental result} and displayed the non-axisymmetry.
}}

Next, we applied a Galilean transformation to the sphere and droplet flow fields. Performing this can enable us to make direct comparisons with the case of uniform flow past a fixed sphere, since this has been reported in literature \cite{taneda1956, johnson1999flow}. The sphere rise velocity is subtracted from each point in the measurement volume, which simulates the situation of a uniform downward flow approaching the sphere. {\color{black}{Figure~\ref{fig:streamlines}} shows the vertical velocity field and streamlines overlaid on it. \color{black}The experimental result is shown in the left figure while the simulation result is shown in the right figure; comparing the left/right figures we can see the streamlines and the wake are very similar, despite the large differences in particle-to-fluid density ratio.}

{\color{black}Lastly, comparisons with prior simulations and experiments of uniform flow past a fixed sphere, at a Reynolds number, Re = 250~\citeauthor{johnson1999flow} \cite{johnson1999flow}, are made. A steady, non-axisymmetric recirculation bubble was also seen in prior work. The length of the recirculation region is found to be marginally shorter than the fixed sphere case in \citeauthor{johnson1999flow}, and our DNS. This can be expected, as our simulations and Ref.\cite{johnson1999flow} deal with a high density particle and a fixed particle, respectively, while our experiment is for a nearly neutrally buoyant sphere. Based on these observations, we believe that the reduced density ratio has the effect of causing a slightly shorter wake for the rising sphere, along with a slightly more pronounced oblique path.}

        \begin{figure}[!h]
    	\centering
    \includegraphics[width = 0.5\textwidth]{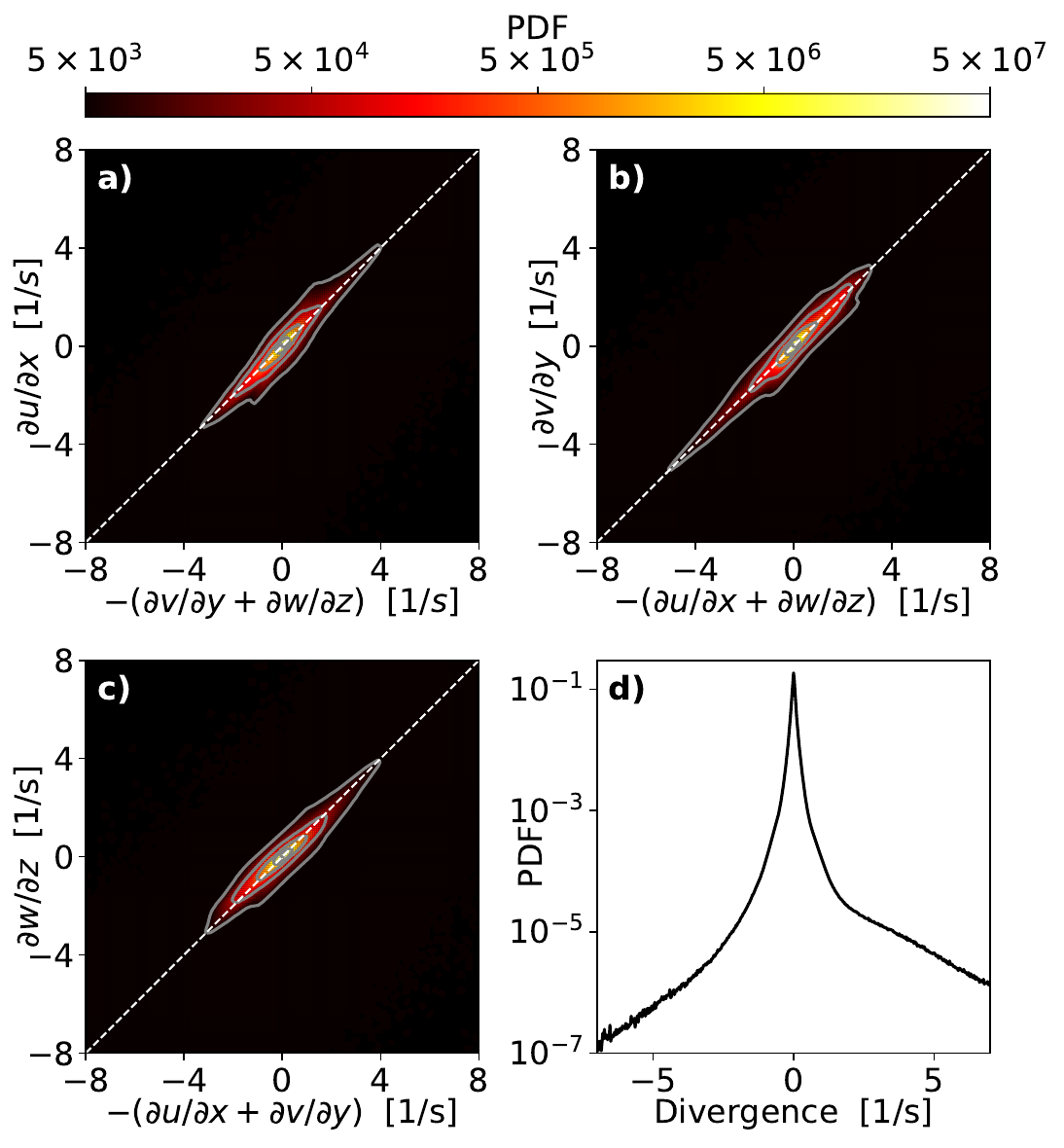}
\caption[Joint PDFs of the divergence for the rising sphere.]{(a)--(c) Joint PDFs showing divergence checks. For each term of the divergence, the distribution against the sum of the two other terms are plotted. For incompressible (divergence-free) flow, the data should fall on the dashed line. The deviations from the dashed line are likely due to errors arising from refraction at the sphere surface. (d) PDF of the divergence on semi-log scale. 
For an estimate of the typical spatial gradients in the flow, $\big \langle\sqrt{\frac{\partial{u}}{\partial{x}}^2+ \frac{\partial{v}}{\partial{y}}^2+ \frac{\partial{w}}{\partial{z}}^2}\big \rangle \approx 2~\text{s}^{-1}$ is found.
The probability of a deviation $P(0.5)\approx 10^{-4} $, this is quite small for typical PIV generated data~\cite{elsinga2006tomographic}.}      \label{fig:divergence_sphere}
\end{figure}

 Following the same procedure in prior sections, divergence checks were performed for the sphere laden flow experiments. The joint PDFs of the three terms of $\nabla \cdot{\bf u}$ 
 are shown in figure~\ref{fig:divergence_sphere}(a--c), and the PDF of $\nabla \cdot{\bf u}$ in  figure~\ref{fig:divergence_sphere}(d).  Overall, the data lie on top of the dashed lines, suggesting that the error in the spatial gradients are minor. 
However, when compared to the case of the single-phase jet at Re = 1000 (figure~\ref{fig:divergence_1000}), some deviations are noticeable in the low probability events. Moreover, the PDF of $\nabla \cdot{\bf u}$ is not symmetric. These deviations, although minor, arise due to refraction effects on the hydrogel sphere surface.
\\\\
           
    \section{Conclusions}

A large-volume, Scanning Stereoscopic Particle Image Velocimetry (SSPIV) system for three-dimensional flow velocity measurements was developed and tested for two different test experiments. The 3D-SSPIV system makes use of a fast-rotating prism to horizontally scan a measurement volume by generating closely-spaced sheets of light. On each of these 2D sheets, stereoscopic PIV was performed, which yielded 3 components of velocity on 2 dimensional sheets~(2D 3C). By combining the 2D 3C fields from the sheets in the scanning direction, all 3 components (3D 3C) of velocity and all 9 components of the spatial velocity gradients were retrieved. Two separate experiments have been performed to gauge the working and the accuracy of the SSPIV system. The first one involved a round jet at a jet Reynolds number Re = 1000. The Re = 1000 jet was found to be stable with a nearly constant spreading angle of 16$^\circ$. This spreading is lower than the 23.6$^\circ$ spreading angle expected for a fully turbulent jet. The velocity profile was shown to be self-similar.  The second experiment performed was on the flow around a freely rising sphere at a Reynolds number Re$_p$ $\approx$ 260. Using the SSPIV system, instantaneous 3D velocity fields were resolved and the wake around the sphere was characterized. The trajectory of the sphere was also tracked, and found to be oblique with a tilt angle of 4$^\circ$ with the vertical. The sphere wake was observed to be steady but non-axisymmetric. A plane comprising the oblique trajectory of the sphere and the vertical line was constructed, and averaged flow fields computed on it. The velocity and vorticity fields revealed the loss of symmetry of the sphere wake, which also explains the oblique rise trajectory. To characterize the flow separation, the sphere rise velocity was subtracted from each grid point in the measurement volume. This yielded velocities in the frame of reference of the rising sphere. {\color{black} Direct numerical simulation of falling droplet at Re$_d$ = 262 was performed and compared with the experimental results. The shape and non-axisymmetry of the wake were similar for the experiment and in the simulation.} The recirculation bubble behind the sphere was characterized, and found to be comparable in  shape to the case of uniform flow past a fixed sphere at Re = 250.  Overall, the 3D-SSPIV method proved useful to measure three-dimensional flow fields in a relatively large measurement volume, whilst being able to resolve all 9 spatial derivatives of velocity.

 The SSPIV system is flexible for use in a variety of applications. The scanning rate and scan volume can be adjusted in three nearly independent ways, namely by varying 1) the size of the prism, 2) the rotation speed of the prism, and 3) the number of faces of the prism. The system can be readily optimized for various applications by only changing the prism, which is a relatively inexpensive component. For different capabilities, the size of the prism and the speed of the rotation can be adjusted. Slightly bigger volumes are limited by the depth of field of the camera-lens combination. Furthermore, by using a precisely programmed pulse generator, the non-linearity in the deflection of the prism could be corrected for to obtain equidistant laser sheets. This can give a homogeneous spatial resolution in the z-direction. 
 
 We thank Sander Huisman, Chao Sun, Detlef Lohse, Kenny Breuer, Gert-Wim Bruggert and Martin Bos for useful discussions and help with design of the system. We thank Xiaojue Zhu for help with video rendering. 


\end{document}